# A New 8/14 Two-Phase Switched Reluctance Motor with Improved Performance

Gholamreza Davarpanah
*Electrical Engineering Department*
*Amirkabir University of Technology*
Tehran, Iran
ghr.davarpanah@aut.ac.ir
https://orcid.org/0000-0002-9794-9049

Hossein Shirzad
*Department of Electrical Engineering*
*Shahid Beheshti University*
Tehran, Iran
h.shirzad@mail.sbu.ac.ir

Jawad Faiz
*Center of Excellence on Applied Electromagnetic Systems, School of Electrical and Computer Engineering*
*University of Tehran* Tehran, Iran
jfaiz@ut.ac.ir

*Abstract*— Despite their simple and robust structure, low cost, and simple cooling system, switched reluctance motors (SRMs) face the challenge of low mean torque. A possible solution is to change the structure of SRMs. This article introduces an innovative combination of the number of rotor teeth and stator teeth of a two-phase switch reluctance motor (TPSRM) with eight teeth for the stator and fourteen teeth for the rotor. As a result of its unique design, which has a short path for passing the main flux, it requires less magnetomotive force. This leads to less core and copper loss, resulting in increased efficiency. Each tooth of the stator in a phase develops a positive torque during the rotation of the rotor, which increases the torque and consequently increases the mean torque of the proposed TPSRM. A current hysteresis control (CHC) is simulated by 2D FEM for the proposed 8/14 TPSRM and the conventional 8/12 TPSRM under the same mechanical load on the shaft to get a current hysteresis reference of 15A at the nominal speed of 600 rpm. To verify the novelty and advantages of the suggested TPSRM, it is compared with the conventional 8/12 TPSRM in terms of mean and peak torque, torque density, and core and copper losses were compared. Lastly, the proposed 8/14 TPSRM is shown to have better performance than the conventional 8/12 TPSRM.

*Keywords— Electric machines, C-core, Switched reluctance motor, finite element method*

## I. Introduction

As the world's largest electricity consumer, electrical machines have always been a hot topic among researchers. With global warming at its peak and the fuel crisis due to recent tensions at its highest point, efficient electrical machines stand out as one of the main ways to offset the repercussions of this precarious situation. Also, the competitive market of home appliances, the heart of which are electrical machines also heats up the situation [1]. In this line of thought, recent studies have been focused on designing electrical machines which can be not only efficient in consuming electricity but also cheap and feasible to produce [2]-[6].

Switched reluctance motor (SRM) are among the candidates which are supposedly able to meet the majority of the mentioned criteria. Of the most sought-after characteristics, the simple structure of SRM is the one that paves the way for producing it at a reasonable cost. In addition to that, given the absence of active excitation parts on the rotor, it is rated as a robust, reliable electrical machine that does not require sophisticated cooling strategies [7]-[9]. With the unprecedented progress of power electronics and effective control strategies, wide speed ranges are also viable in the SRMs. Despite benefitting from these inherent merits, SRMs suffer from a major drawback: low mean torque. Low torque density and a noisy operation are also listed as the potential demerits of SRMs. Given the overriding need for an efficient SRM capable of producing enough mean torque while having a sufficient torque density, setting the stage for optimizing it so as to overcome the mentioned drawbacks is inevitable, as is the case with other types of electrical machines [10]. In this regard, numerous strategies have been introduced by different authors [11]-[12].

An effective method of enhancing the mean torque and torque density as well as minimizing the torque ripple is to come up with a novel topology and optimize its structure [13]. One of the things that makes SRMs so attractive to the industry is that they can be designed in different ways, and this can be accomplished by increasing the number of poles on the stator and rotor. A three-phase SRM employing C-core is introduced in [14] which outperforms conventional SRMs [15], [16]. The adjacent teeth of the stator of a C-core are taken into account, thereby forming one path for the generated flux. As the rotor rotates, a C-core stator tooth develops a negative torque, while its other tooth develops a positive torque. As a result, the electromagnetic torque is reduced in aggregate. [17] presents a novel structure in which the negative torque problem was effectively addressed. In fact, mean torque, torque density, and other major characteristics were enhanced given the absence of negative torque, reduced flux reversal effects, and shortened flux paths. Different modular structures are also presented in [18] that employ both segmented and non-segmented rotors. It was concluded that both motors generate higher average torque than conventional SRM while consuming less steel. A novel configuration has been proposed in [19], in which the number of stator teeth is higher than that of the rotor teeth. At a stator excitation current of 3 A, the proposed motor's mean torque increases by 24%, but its torque ripple also increases by 18% when compared to a conventional SRM.

Recently, more attention has been paid to TPRSMs and they are studied as a potential candidate to be used in home appliances [20], industrial tools [21] and also electrical vehicles [22]-[25]. This is due to the fact that the drive circuits of TPSRMs have fewer power electronics components than those of SRMs with three phases. In fact, TPSRMs have fewer switches and diodes, resulting in lower converter prices and a lower system cost [8], [9], and [26].

A new combination of the number of rotor teeth and stator teeth is presented in this article for a two-phase switch reluctance motor (TPSRM). As the stator is formed of C-cores, the magnetic flux paths are shorter, resulting in lower core losses. A lower magneto-motive force (MMF) and consequently a smaller stator current reduces copper losses as well. During the rotor's movement, every four teeth of the stator develop a positive torque, which increases the torque and, consequently, increases the mean torque of the proposed TPSRM. Finally, a conventional 8/14 TPSRM is compared with the proposed topology to validate its performance.

This paper is structured as follows: Section II describes the topologies of the proposed SRM and the one with which it is to be compared. Section III introduces the optimization procedure. As part of the 2D finite element analysis (FEA), flux density analysis is provided for aligned, half-aligned, and unaligned conditions. A comparison is presented in Section IV between the results obtained for the proposed SRM and those obtained for the conventional SRM. Both static and dynamic electromagnetic performances are put side by side to ensure a fair comparison. A two-dimensional finite element method (FEM) is adopted to obtain the electromagnetic characteristics of the two TPSRMs. A conclusion is provided in Section V.

## II. Motor Topology

The base concept of 8/14 Two-phase SRM (TPSRM) with stator poles of C-core form is proposed here. This novel topology and its fundamental principles using two pairs of C-core for the stator are presented.

### A. Concept

As illustrated in Fig. 1(a), there are two C-cores for each phase, and to eliminate the axial forces exerted on the stator and to maintain the symmetry of forces applied to it, the C-cores are placed opposite each other. For two-phase SRM, four C-cores are required so that magnetic flux passing through one tooth of the C-core and entering the air gap passes through the next tooth of the same C-core (Fig. 1(b)). Therefore, two teeth of the rotor must be aligned with two stator C-core teeth. The rotor pole pitch is as follows:

$$\theta_{rp} = 360°/N_r \quad (1)$$

where $N_r$ is the number of rotor poles. So, according to Fig. 1(c), the angle between the two teeth of the C-core that in this design has been considered $2\theta_{rp}$ is as follows:

$$\alpha = 2\theta_{rp} = 720°/N_r \quad (2)$$

since the number of C-cores for a two-phase SRM is four, the angle from the center of one tooth of the C-core and the center of the adjacent tooth of another c-core is γ. According to Fig. 1(c), in this design, one phase is aligned, and the other phase must be unaligned. Therefore γ has been considered as follows:

$$\gamma = 1.5\theta_{rp} = 540°/N_r \quad (3)$$

since the considered SRM is a two-phase SRM and has two C-core pairs, α and γ are as follow:

$$4(\alpha + \gamma) = 360° \Rightarrow (\alpha + \gamma) = 90° \quad (4)$$

the number of rotor teeth can be easily obtained by putting equations (2) and (3) in equation (4), which is 14. By arriving at the number of rotor teeth, a structure with a unique number of rotor and stator teeth for a two-phase SRM is obtained (8/12 two-phase SRM). In addition to that, it is noted that the main magnetic flux path in this structure is short, resulting in reduced core losses, which increases the SRM's efficiency.

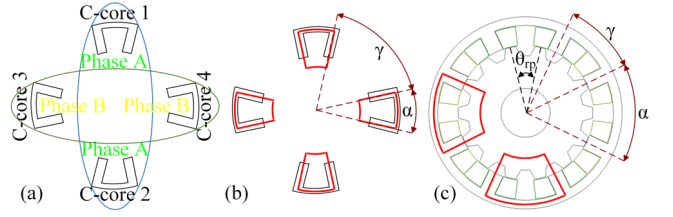

Fig. 1. Concept of the proposed 8/14 TPSRM.

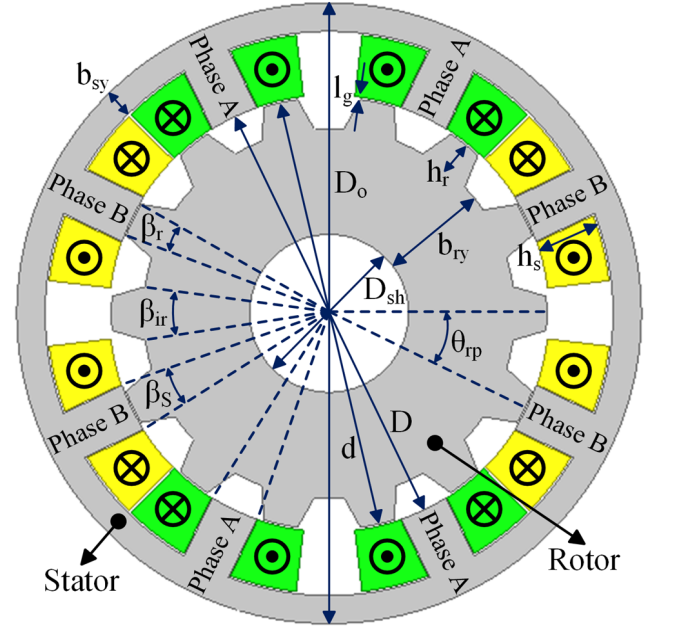

Fig. 2. Geometry of the proposed 8/14 TPSRM.

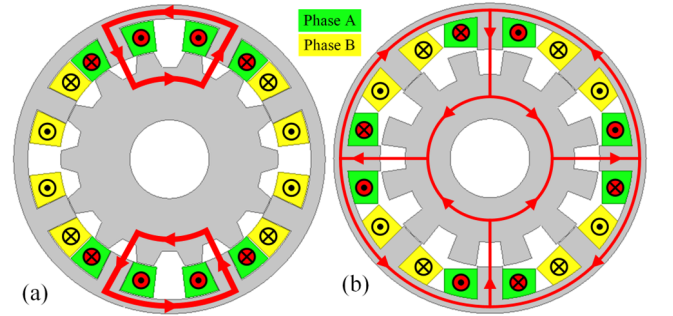

Fig. 3. Topology and main flux paths of the (a) proposed 8/14 and (b) conventional 8/12 TPSRMs.

### B. Motor Topologies

Fig. 2 shows the 2-D cross-section of the proposed 8/14 TPSRM structure. There are two parts to the proposed TPSRM: fixed parameters and variable parameters. Fig. 3 shows the proposed 8/14 TPSRM topology and conventional 8/12 TPSRM topology that is intended for comparison. Conventional 8/12 TPSRM is an acceptable candidate for comparison since it has the closest combination of rotor and stator teeth with the proposed 8/14 TPSRM. The proposed and conventional SRMs are designed for a standard application with equal volumes (the same stack length of the stator and the same outer diameter of the stator). Also, the same air gap length and the same shaft diameter. The main flux paths of the proposed and the conventional SRMs are given shown in Fig. 3. Table I presents the constant and key parameters of the dimensions of two TPSRMS. Here, this parameter and nominal current hysteresis is considered based on the volume of the SRMs, available winding space, the diameter of copper, and the power supply used.

TABLE II
OPTIMIZED PARAMETERS AND DIMENSIONS OF THE TWO TPSRMs.

| Parameter | Region of optimization variables | 8/14 TPSRM | Region of optimization variables | 8/12 TPSRM |
|---|---|---|---|---|
| Stator yoke thickness, $b_{sy}$ (mm) | $3.48 \leq b_{sy} \leq 6.96$ | 5.18 | $4.22 \leq b_{sy} \leq 8.44$ | 5.00 |
| Stator pole length, $h_s$ (mm) | $9.44 \leq h_s \leq 13.21$ | 11.8 | $9.58 \leq h_s \leq 13.41$ | 11.98 |
| Stator pole arc, $\beta_s$ (deg) | $7.71 \leq \beta_s \leq 18.00$ | 12.85 | $9.00 \leq \beta_s \leq 21.00$ | 13.3 |
| Rotor pole arc, $\beta_r$ (deg) | $7.71 \leq \beta_r \leq 18.00$ | 9.07 | $9.00 \leq \beta_r \leq 21.00$ | 15.30 |
| Inner Rotor pole arc, $\beta_{ir}$ (deg) | $9.00 \leq \beta_{ir} \leq 21.00$ | 16.64 | $9.00 \leq \beta_{ir} \leq 21.00$ | 15.84 |
| Rotor pole length, $h_r$ (mm) | $4.45 \leq h_r \leq 8.91$ | 5.79 | $4.45 \leq h_r \leq 8.91$ | 8.89 |

TABLE I
CONSTANT KEY DIMENSIONS OF TWO TPSRMs.

| Parameter | 8/14 TPSRM | 8/12 TPSRM |
|---|---|---|
| Stator outer diameter, $D_o$ (mm) | 114 | 114 |
| Air-gap length, $l_g$ (mm) | 0.3 | 0.3 |
| Stator inner diameter, $D$ (mm) | 80.04 | 80.04 |
| Rotor outer diameter, $d$ (mm) | 79.44 | 79.44 |
| Shaft diameter, $D_{sh}$ (mm) | 29 | 29 |
| Stack length, $L$ (mm) | 40 | 40 |
| Nominal current hysteresis $I_{hys}$ (A) | 15 | 15 |
| Number of turns per pole, $T_{pole}$ | 94 | 94 |

## III. DETERMINATION OF OPTIMAL DESIGN PARAMETERS

As seen in Fig. 2, numerous parameters must be optimized to achieve better performance from the proposed 8/12 TPSRM. The finite element method (FEM) is time-consuming to optimize when many variables are involved because of the high computation. In some papers, FEM sensitivity analysis is used to investigate the impact of parameters on the cost function. The main problem is that the affection of parameters is individually considered, while they might affect other parameters.

### A. Design and Optimization of Motor

This section discusses the optimization of the proposed two-phase 8/14 SRM and conventional two-phase 8/12 SRM. A critical step in the design procedure of the two-phase SRM is to find the highest mean torque. For optimizing parameters, sensitivity analysis, which is a method of changing one parameter and keeping the others constant, is unsuitable for obtaining optimal parameters. Therefore, the genetic algorithm (GA) method is brought into the limelight. The GA is considered to get both SRMs' final dimensions. The purpose of this optimization is to improve the mean torque. The analytical model of torque calculation is presented in [27]. Therefore, the function used to increase the mean electromagnetic torque is defined as follows:

$$T_{mean} = \frac{1}{T}\int_0^T T_e(t)dt \quad (5)$$

The region of each variable, which is presented based on equations in [26], and the final optimum dimensions of the parameters of the two SRMs are listed in table II. For additional investigation, the proposed and conventional SRMs with their optimized dimensions are analyzed through the FEM.

### B. Predicted Flux Analysis

Fig. 4(a) and (b) illustrate the flux paths within the proposed 8/12 TPSRM for the unaligned condition under the excitation of phase A and the aligned condition under the excitation of phase B, respectively. The bold red line indicates the main flux path and its direction going through the adjacent stator poles of the same phase, while the thin red line denotes another path of flux having a minimal amount of whole magnetic flux. One of the main characteristics of the present paper is suggesting an SRM with a short path for passing the main flux of the stator phase teeth, which leads to less core loss. Also, both stator teeth in the C-core along the rotor rotation develop positive torque leading to higher mean torque and increased torque density in the proposed SRM. Finally, it can contribute to higher efficiency.

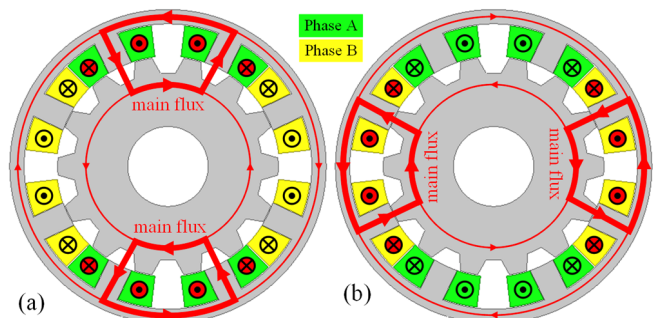

Fig. 4. Main flux paths within the proposed 8/14 TPSRM when (a) phase A is excited under the unaligned condition and (b) phase B is excited under the aligned condition.

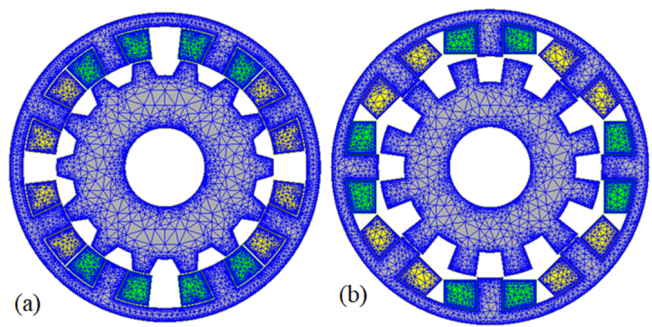

Fig. 5. Meshed model of the (a) final proposed 8/14 TPSRM, and (b) 8/12 conventional TPSRM.

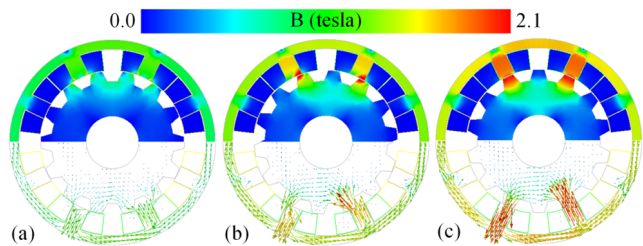

Fig. 6. Magnetic flux density distribution and magnetic flux density vectors in the proposed 8/14 TPSRM under (a) unaligned, (b) half-aligned, and (c) aligned conditions at the nominal 15A excited current.

### C. Flux Density Analysis

The mesh models used for both proposed TPSRM and conventional TPSRM for 2D analysis are shown in Fig. 5. These mesh elements are small enough to achieve satisfactory accuracy. In Fig. 6, magnetic flux density distribution and magnetic flux density vectors are presented in the proposed 8/14 TPSRM under unaligned, half-aligned, and aligned conditions at the nominal 15A excited current. The magnetic flux density distribution and magnetic flux density vectors of different parts of the rotor and stator at three conditions of the rotor show that the magnetic flux density in the stator yoke and rotor yoke is lower than that of the stator and rotor poles. The maximum magnetic flux density exists in the unaligned position in the stator tooth and the aligned position in the rotor tooth. Fig. 7 illustrates magnetic flux paths in the proposed 8/14 TPSRM under unaligned, half-aligned, and aligned conditions at the nominal 15A excitation current. As has been predicted, the majority of the magnetic flux pass through the yoke between the poles of the excited phase, shortening the path of the main part of the magnetic fluxes. Also, in this structure, there is no flux reversal in the stator core. This unique topology reduces the core losses of a typical TPSRM, and all these factors lead to the improvement of the efficiency of the proposed SRM.

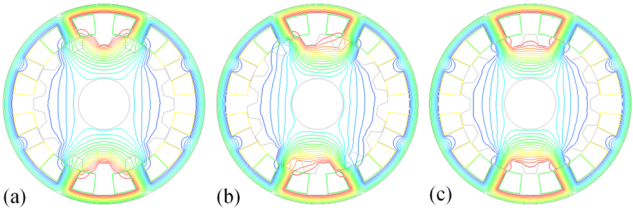

Fig. 7. Magnetic flux paths in the proposed 8/14 TPSRM under (a) unaligned, (b) half aligned, and (c) aligned conditions at the nominal 15A excited current.

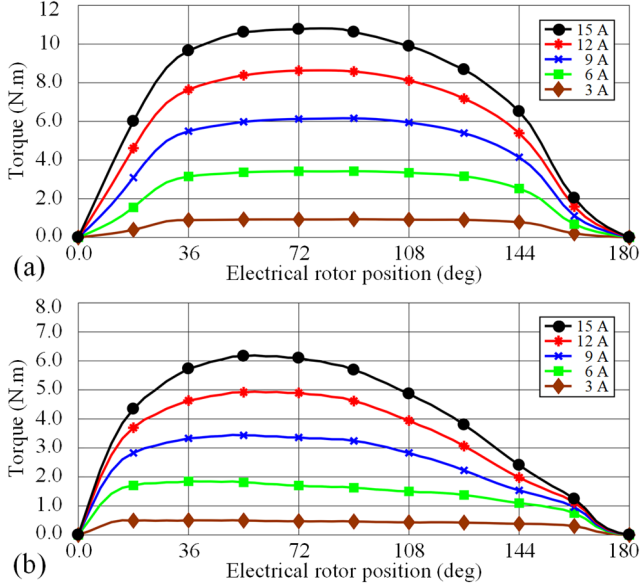

Fig. 8. Static torque profile of (a) the proposed 8/14 TPSRM and (b) the conventional 8/12 TPSRM for different currents.

## IV. SIMULATION RESULT AND COMPARISON

This section presents the finite element method (FEM) and a comparison of the proposed 8/14 TPSRM with conventional 8/12 TPSRM. A two-dimensional finite element method (FEM) is adopted to obtain the electromagnetic characteristics of the two TPSRMs.

### A. Static Simulation Results

Fig. 8 presents the static torque profiles of the proposed 8/14 TPSRM and the conventional 8/12 TPSRM after optimization by GA at five different levels of the excitation current and rotor rotation (in electrical degree) from the unaligned position to the aligned position (between 0 to 180 electrical degrees). Since the changes in reluctance cause the generation of torque and when the rotor rotates from the unaligned position to the aligned position, the reluctance changes are variable, and these reluctance changes are not symmetrical, so the torque profile obtained in Fig. 8 is also not symmetrical. Fig. 9 presents the average and peak torques of the two TPSRMs for different stator currents, from which a significant increase for the proposed 8/14 TPSRM in both quantities can be observed.

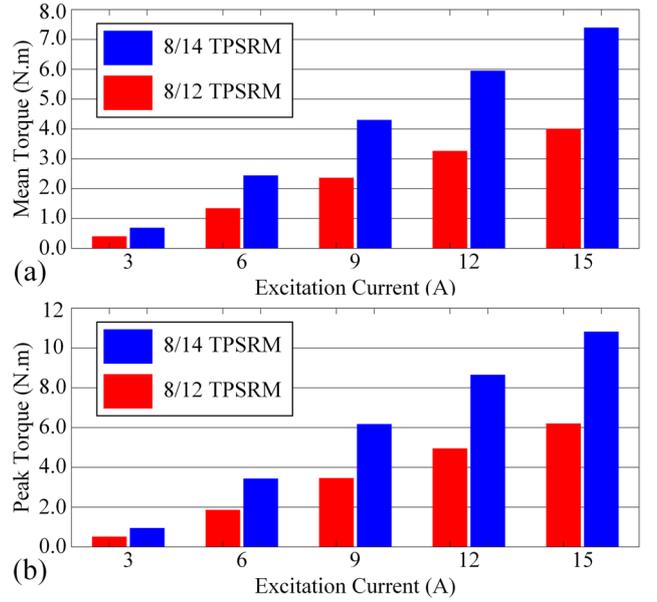

Fig. 9. (a) mean and (b) peak torque of the proposed 8/14 TPSRM and the conventional 8/12 TPSRM for different currents.

### B. Dynamic Simulation Results

A current hysteresis control (CHC) is simulated by 2D-FEM for the proposed 8/14 TPSRM and the conventional 8/12 TPSRM under the same mechanical load on the shaft to get a current hysteresis reference of 15A at the nominal speed of 600 rpm. An asymmetric bridge converter is used to drive the two TPSRMs. The turn-on period of each phase is 180 electrical degrees. The hysteresis band δ is set to 0.4 A (14.8 A to 15.2 A). That means when the current of phase is greater than 15.2 A, the switches of that phase turn off, and the current of phase decreases until the current of phase becomes lower than 14.8 A, then the switches of that phase turn on; this cycle continues for 180 electrical degrees. The voltage of the DC link is 150 V. Steady-state voltage, current hysteresis, flux linkage, and torque waveforms for the final proposed 8/12 TPSRM and the conventional 8/12 conventional TPSRM at the current hysteresis of 15 A, 600 rpm, and CHC mode are given in Fig. 10.

### C. Comparison with Conventional SRMs

Here are static torque densities comparisons between the aforementioned TPSRMs, given in Fig. 11. It is observed that a remarkable increase in the torque density has been attained in the proposed 8/14 TPSRM. Table III shows the static mean and peak torque for both proposed 8/14 and conventional 8/12 TPSRMs, as well as the increase of the values mentioned above for the proposed 8/14 TPSRM in comparison with the conventional 8/12 TPSRM for same phase excitation. Also, the proposed 8/14 TPSRM has a higher maximum torque. At 15 A, the static mean torque of the proposed 8/14 TPSRM is

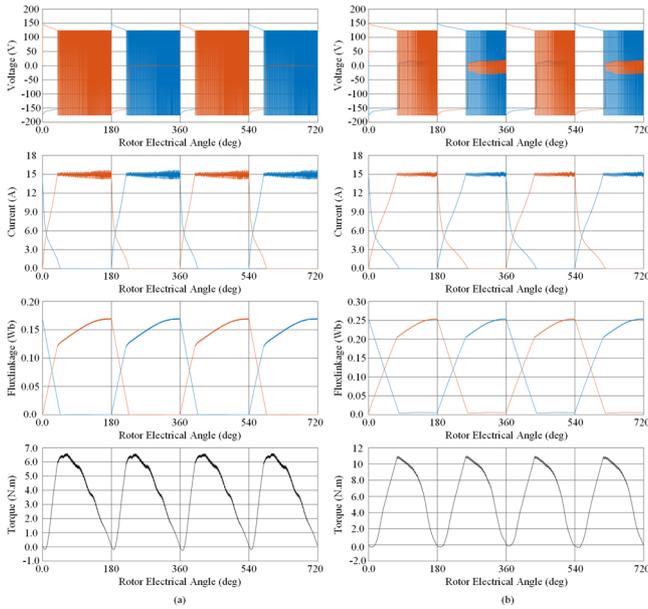

Fig. 10. Steady-state voltage, current hysteresis, flux linkage, and torque waveforms for (a) the conventional 8/12 TPSRM and (b) the final proposed 8/14 TPSRM at the current hysteresis of 15 A, 600 rpm, and CHC mode.

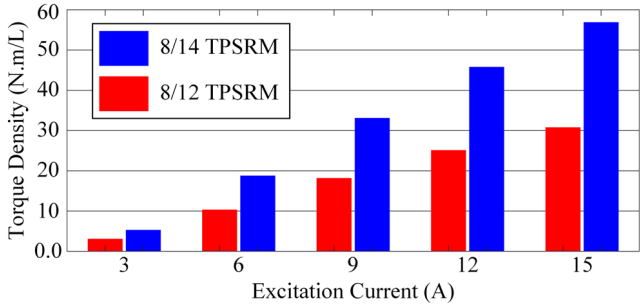

Fig. 11. Comparison of torque density between the proposed 8/14 TPSRM and the conventional 8/12 TP SRM for different currents.

TABLE III
COMPARISON OF STATIC SIMULATED MEAN AND PEAK TORQUE FOR DIFFERENT CURRENT LEVELS.

| Phase current (A) | Mean torque (N.m) 8/14 TPSRM | Mean torque (N.m) 8/12 TPSRM | Peak Torque (N.m) 8/14 TPSRM | Peak Torque (N.m) 8/12 TPSRM | 8/14 TPSRM compared to 8/12 TPSRM Mean torque increase (%) | 8/14 TPSRM compared to 8/12 TPSRM Peak torque increase (%) |
|---|---|---|---|---|---|---|
| 3 | 0.682 | 0.394 | 0.938 | 0.505 | 73.09 | 85.74 |
| 6 | 2.435 | 1.335 | 3.423 | 1.849 | 82.39 | 85.12 |
| 9 | 4.295 | 2.357 | 6.163 | 3.448 | 82.22 | 78.74 |
| 12 | 5.945 | 3.257 | 8.641 | 4.940 | 82.52 | 74.91 |
| 15 | 7.388 | 3.993 | 10.808 | 6.189 | 85.02 | 74.63 |

85.02% higher. Since both TPSRMs have the same volume, the advantage and novelty of the C-core connected are confirmed here. The details of the predicted dynamic behavior of two TPSRMs at 600 rpm with current hysteresis control (CHC) mode, torque ripple, and utilized iron are presented in table IV. It is seen that the proposed 8/14 TPSRM has a better performance than the conventional 8/12 TPSRM. Also, the mean torque and the overall efficiency of the proposed TPSRM are observed to be larger than conventional TPSRM. The mean torque and efficiency of the proposed TPSRM are 55.85% and 5.43% higher than the conventional TPSRM.

TABLE IV
PREDICTED DYNAMIC BEHAVIOR OF TWO TPSRMs AT 600 rpm WITH THE CURRENT HYSTERESIS CONTROL (CHC) MODE.

| Parameter | 8/14 TPSRM | 8/12 TPSRM |
|---|---|---|
| Motor volume (mL) | 129.96 | 129.96 |
| Nominal speed, $N_r$ (rpm) | 600 | 600 |
| RMS phase current $I$ (A) | 9.36 | 9.98 |
| Mean torque $T_{ave}$ (N.m) | 5.86 | 3.76 |
| Torque ripple (%) | 150.65 | 161.37 |
| Iron weight (Kg) | 1.905 | 1.820 |
| Output power $P_d$ (W) | 368.19 | 236.24 |
| Stator copper losses $P_{cu}$ (W) | 71.19 | 50.80 |
| Total core loss $P_c$ (W) | 4.82 | 13.45 |
| Input power $P_{in}$ (W) | 444.2 | 300.49 |
| Torque per ampere (Nm/A) | 0.626 | 0.376 |
| Power per ampere (W/A) | 39.37 | 23.67 |
| Torque density (N.m/L) | 45.09 | 20.93 |
| Efficiency $k_e$ (%) | 82.89 | 78.62 |

## V. CONCLUSION

This paper presented and analyzed a novel configuration for an 8/14 two-phase SRM (TPSRM) using FEM. The principal characteristics of this TPSRM, including mean and peak torque, torque density, magnetic flux lines, flux linkage, and magnetic flux density, were obtained. Analysis and a comprehensive comparison between a proposed 8/14 TPSRM and the conventional 8/12 TPSRM in which stator outer diameter, stack length, rotor outer diameter, air gap, and shaft diameter are equal were carried out to show the merits of the proposed 8/14 TPSRM. As the magnetic flux paths are shorter in the proposed 8/14 TPSRM, the core losses are reduced; on the other hand, the required magneto-motive force and, consequently, the stator current is smaller, resulting in reduced copper losses. Also, the flux second path leads to a reduction in the reluctance of the flux path and, thus, an increase in the torque. The dynamic behavior of the two TPSRMs was analyzed under the current hysteresis control (CHC) mode. The results showed that at a current hysteresis of 15 A, for similar conditions with a constant volume, the proposed TPSRM develops 55.85% more mean torque compared with the conventional TPSRM. In addition, its efficiency is 5.43% higher. These comparisons showed the merits of the proposed TPSRM, and the superiority of the proposed TPSRM is proven in comparison with the conventional TPSRM.